\documentclass[twocolumn,showpacs, secnumarabic,amssymb, nobibnotes, aps, prd]{revtex4}

\usepackage{graphicx}
\usepackage{dcolumn}
\usepackage{bm}

\begin{document}

\title{Tuning magnetic avalanches in Mn$_{12}$-ac}

\author{S. McHugh}
\author{Bo Wen}
\author{Xiang Ma}
\author{M. P.  Sarachik}
\email{sarachik@sci.ccny.cuny.edu}

\affiliation{
Department of Physics\\ City College of New York, CUNY\\
New York, New York 10031, USA}

\author{Y. Myasoedov}
\author{E. Zeldov}
\affiliation{
Department of Condensed Matter Physics\\
The Weizmann Institute of Science\\
Rehovot 76100, Israel}

\author{R. Bagai}
\author{G. Christou}
\affiliation{
Department of Chemistry\\
University of Florida\\
 Gainesville, Florida 32611, USA}
 
\begin{abstract}

Using micron-sized Hall sensor arrays to obtain time-resolved measurements of the local magnetization, we report a systematic study in the molecular magnet Mn$_{12}$-acetate of magnetic avalanches controllably triggered in different fixed external magnetic fields and for different values of the initial magnetization.  The speeds of propagation of the spin-reversal fronts are in good overall agreement with the theory of magnetic deflagration of Garanin and Chudnovsky  \cite{Garanin}.

\end{abstract}

\pacs{75.45.+j, 75.40.Gb, 47.70.Pq}

\maketitle

\section{\label{sec:level1}Introduction}
Mn$_{12}$-acetate is a crystal composed of magnetic molecules, each of which behaves as a high-spin, high-anisotropy magnet \cite{Lis, Sessoli}.   At low temperatures, the 12 Mn atoms are strongly coupled via superexchange to form a ferrimagnet with a net (rigid) spin $S=10$.  Magnetic interactions between the molecules are thought to be negligible so that  Mn$_{12}$ can be modeled by an effective spin Hamiltonian:
\begin{eqnarray}
\mathcal{H} = -DS_z^2 - AS_z^4 -g\mu_BS_zB_z + \mathcal{H}_\perp,
\label{Hamiltonian}
\end{eqnarray}
where $B_z$ is a magnetic field applied along the $c$-axis of the crystal, $S_z$ is the $z$ component of the spin, $D = 0.548$ K, $A = 1.17 \times 10^{-3}$ K, $g = 1.94$, and $\mathcal{H}_\perp$ represents small symmetry-breaking terms that allow tunneling  across the anisotropy barrier \cite{g Factor, Friedman, Hamiltonian}.  The energy barrier against magnetic reversal, $U$, is easily calculated from Eq. \ref{Hamiltonian}.  The magnetic relaxation rate for an individual molecule becomes sufficiently slow at low temperatures ($< 2$ K) that the magnetization of the crystal can be prepared and maintained in a metastable state for time periods well in excess of the experimental times.  Once in this metastable state, the magnetization may relax as an abrupt ($< 1$ms) ``magnetic avalanche,'' an exothermic process involving the release of Zeeman energy \cite{Paulsen}.  These spatially inhomogeneous reversals proceed as a traveling ``front'' between regions in the crystal with opposing magnetization and have been described as magnetic deflagration in analogy with chemical deflagration \cite{Suzuki, Garanin}.  

In general, the speed of a deflagration front is governed by two parameters: the thermal diffusivity, $\kappa$, which specifies the rate at which heat diffuses through the medium, and the reaction rate of the constituents, $\Gamma(T_f)$, where  $T_f$ is the ``flame temperature'' produced by the reactants near the front.  Combining these parameters gives an approximate expression for the speed, $v \sim \sqrt{\kappa\Gamma}$ \cite{Landau}.  In the case of magnetic deflagration, the medium through which heat flows is the crystal, and $\Gamma$ is the relaxation rate of the metastable spins, which obeys an Arrhenius law,
\begin{eqnarray}
\Gamma = \Gamma_0 \mbox{ exp}\left[-U(B)/T\right],
\label{Arrhenius}
\end{eqnarray}
with $\Gamma_0 = 3.6\times 10^7$ s$^{-1}$ \cite{Gamma_0, Gomes}.  The relaxation rate can be increased by lowering the barrier with an increasing external magnetic field or by increasing the temperature, $T$.  Although $\kappa$ has not been measured for Mn$_{12}$-ac, a value of $\kappa \sim 10^{-5}$ m$^2$/s was deduced from the avalanche data in Ref. \cite{Alberto2}.  Suzuki et al. \cite{Suzuki} showed that the speeds of magnetic avalanches can be modeled approximately as, $v\sim \sqrt{\kappa\Gamma_0}$ exp$[-U/(2T_f)]$, where $T_f$ is the temperature at or near the propagating front associated with the energy released by the reversing spins.  The theory of magnetic deflagration stands in qualitative agreement with experiments, yet more precise quantitative confirmation remains an open experimental challenge \cite{Suzuki, Alberto1, Alberto2, Garanin, McHugh}, which is undertaken here. 

There are two parameters under experimental control when performing magnetic avalanche studies on Mn$_{12}$-ac: the external magnetic field and the initial magnetization which tunes the metastable spin density and thus the available ``fuel".  Varying the external magnetic field affects both the barrier against spin reversal and the energy released which determines the temperature $T_f$ produced by the reversing spins.  The variation of the metastable spin density affects primarily the energy released and therefore, $T_f$.  By varying these parameters independently, we are able to explore the wide ranging conditions in which avalanches may be triggered.  In particular, we report systematic studies for three classes of avalanche preparations.  The first class (I) contains avalanches triggered at various external fields with fixed (maximum) initial magnetization.  For these avalanches, both $U$ and $T_f$ vary.  The second class (II) contains avalanches triggered at a fixed external field, but with various initial magnetizations.  Avalanches of this class differ primarily through $T_f$, with $U$ varying only through the internal fields.  And the final class (III) contains avalanches triggered with various initial magnetizations and external fields, such that $T_f$ is approximately constant while $U$ varies (details below).

We report the speeds of propagation of the fronts for these three classes of avalanches, allowing us to make a thorough comparison with the theory of magnetic deflagration.  Our results are in overall agreement with the theory.  With certain simplifying assumptions detailed below, we obtain temperatures between $5$ and $16$ K and thermal diffusivities ranging from $\kappa = 1.2\times10^{-5}$ to $8.7\times10^{-5}$ m$^2$/s, consistent with the value of $\kappa$ estimated in Ref. \cite{Alberto2}. 

\begin{figure}[htbp]
\begin{center}
  \includegraphics[width=2.75in]{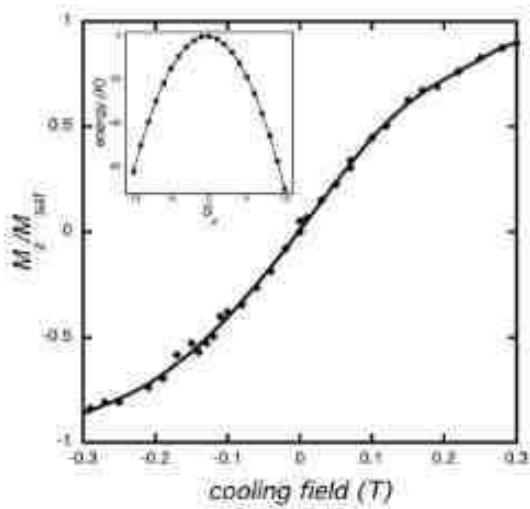}
\caption{The initial magnetization, $M_z$, as a function of applied cooling field.  $M_{sat}$ is defined as the magnetization with all the spins aligned in the positive direction.  The change of magnetization during an avalanche is $\Delta M = M_{sat} - M_z$.  The same curve is found for all crystals measured.  The inset shows the energy levels biased with a +0.3 T field.}
\label{MvsH}
\end{center}
\end{figure}

\section{\label{sec:level1}Experiment}
The magnetization dynamics were studied with an array of $30\times 30$ $\mu$m$^2$ Hall sensors spaced apart by $80$ $\mu$m center-to-center \cite{Avraham}.  Crystals of Mn$_{12}$-acetate with dimensions about $1.0\times0.2\times0.2$ mm$^3$ were attached to the array with Apiezon M grease.  The crystal was encased in grease along with a constantan wire placed near the sample for use as a heater in an arrangement similar to that used in Ref. \cite{McHugh}.  The entire assembly was immersed in $^3$He at temperatures down to $< 300$ mK.  

Prior to triggering an avalanche, the sample was prepared in a metastable magnetic state.  To do this, the sample was cooled from a high temperature ($\approx 6$ K) down to $300$ mK in the presence of a small external magnetic ``cooling" field between $\pm 0.3$ T.  The inset of Fig. \ref{MvsH} is a schematic of the energy levels biased with a +0.3 T cooling field; as shown in Fig. \ref{MvsH}, the magnetization of the sample, $M_z$, depends on the magnetic field in which it was cooled.  At 0.3 K, only the $S_z = \pm 10$ states are appreciably occupied.  $M_z$ reflects the ratio of these occupied states.  Once the sample is well below the blocking temperature, the external field can be changed without changing the magnetization (the system is $blocked$).  The field is then increased to a predetermined value ($\geq +1.25$ T) \cite{minor complication}.  When the field has stabilized ($\sim 1$ min), a current is passed through the wire heater gradually raising the temperature and triggering the avalanche.  For more details on triggering avalanches with this method, see Ref. \cite{McHugh}.

All avalanches reported here were triggered in a positive field.  The amount of metastable magnetization that reverses during an avalanche is given by $\Delta M = |M_{sat}-M_z|$, where $M_z$ is the initial magnetization and $M_{sat}$ is the magnetization with all spins aligned in the positive direction.  For full magnetization reversal, $\Delta M = 2M_{sat}$.  For convenience, we introduce the parameter $\Delta M/2M_{sat}$ as a dimensionless measure of the initial metastable magnetization density.  As an example, cooling the sample in zero field leads to $M_z = 0$, or $\Delta M/2M_{sat} = 0.5$.  

\begin{figure}[htbp]
\begin{center}
  \includegraphics[width=3in]{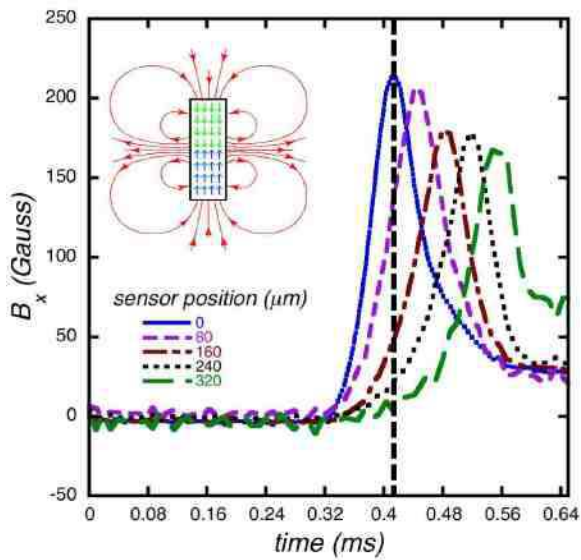}
\caption{The time response of five equally spaced Hall sensors placed along the length of the crystal.  The peak indicates the arrival of the magnetization interface (avalanche front).  The inset shows the traveling front that separates spin-up and spin-down regions.  Note the transverse magnetic field lines at the position of the spin-reversal front.}
\label{scopeData}
\end{center}
\end{figure}

We prepared the three classes of avalanches by varying the initial magnetization, $M_z$, and the external magnetic field $H_z$.  Variation of $H_z$ tunes the barrier, $U$, as well as the average energy released per molecule during an avalanche,
\begin{eqnarray}
\langle E \rangle  =  2 g\mu_B S B_z \left(\frac{\Delta M}{2M_{sat}}\right),
\label{E}
\end{eqnarray}
where $B_z = \mu_0(H_z + M_z)$.  Avalanches of class I are those with initial magnetization $M_z = -M_{sat}$, i.e., $\Delta M/2M_{sat} = 1$, and $H_z$ is varied.  Class II avalanches are triggered at a fixed external field, with various initial magnetizations.  Finally, class III avalanches are triggered at various $H_z$ and $M_z$, such that $\langle E\rangle$ remains constant.

We collected avalanche data on four different crystals with dimensions $1.00\times0.20\times0.20$ mm$^3$ (Sample A), $1.20\times0.10\times0.10$ mm$^3$ (Sample B),  $0.80\times0.15\times0.15$ mm$^3$ (Sample C), and $1.00\times0.25\times0.25$ mm$^3$ (Sample D).  We report detailed data on crystal A.  Although the absolute values of the avalanche speeds differed for different crystals (as discussed in detail later in this paper), similar behavior was obtained for all crystals as a function of the experimental parameters.

\section{\label{sec:level1}Results}

As was shown by Suzuki et al. \cite{Suzuki}, the avalanche progresses through the crystal in a similar fashion to that of a domain wall in a ferromagnet \cite{!domain wall}.  There is an interface separating regions of opposing magnetization, which produces a large transverse magnetic field, $B_x$ near the front, as shown schematically in the inset of Fig. \ref{scopeData}.  Figure \ref{scopeData} shows the time response of five equally spaced Hall sensors produced by a zero field cooled avalanche.  At time $t = 0$, the magnetization is zero.  At $t\approx 0.32$ ms, the magnetization begins to reverse near the first sensor as the avalanche front approaches.  At $t=0.41$ ms, the signal on the first sensor is maximum indicating the avalanche arrival at position $0$ $\mu$m.  By $t=0.65$ ms all spins have reversed and the sample is completely magnetized.  The avalanche speed is deduced from the arrival time of the peak at each sensor and the known spacing between the sensors.

\begin{figure}[htbp]
\begin{center}
  \includegraphics[width=3in]{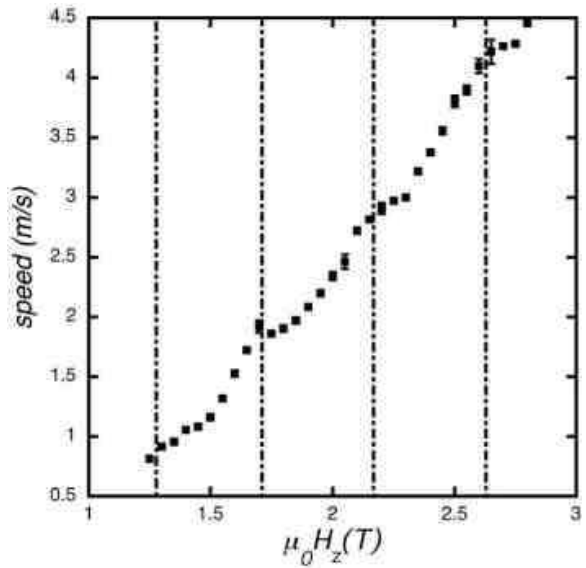}
\caption{Propagation speed as a function of applied magnetic field for Class I avalanches in crystal A for Class I avalanches, $\Delta M/2M_{sat} = 1$}
\label{fixedM}
\end{center}
\end{figure}

\begin{figure}[htbp]
\begin{center}
  \includegraphics[width=3in]{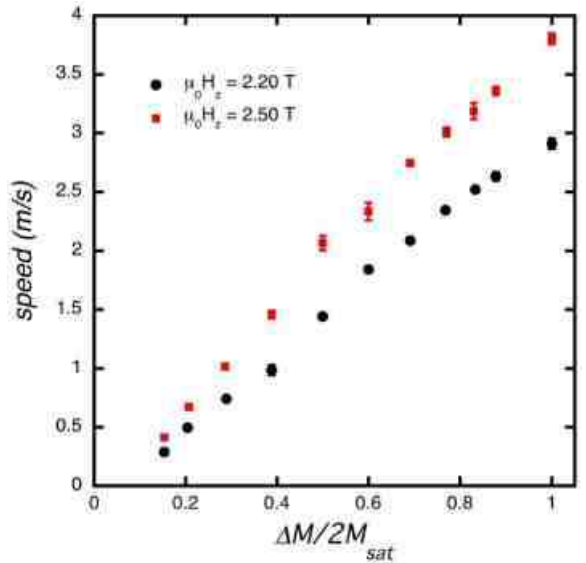}
\caption{Propagation speed as a function of $\Delta M/2M_{sat}$ for Class II avalanches in crystal A.  The field was fixed at $2.50$ T and $2.20$ T, while $\Delta M/2M_{sat}$ was varied between about $0.10$ and $1.00$.}
\label{fixedU}
\end{center}
\end{figure}

\begin{figure}[htbp]
\begin{center}
  \includegraphics[width=3in]{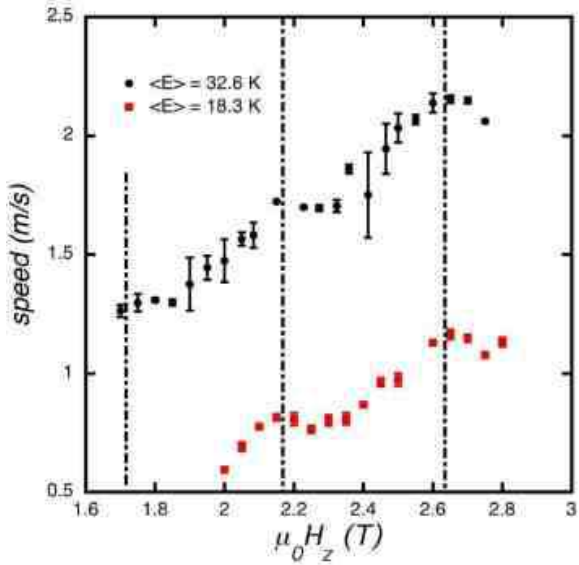}
\caption{Propagation speed as a function of applied magnetic field for Class III avalanches in crystal A.  The initial magnetization and external field are adjusted to hold the average energy released per molecule constant at $\langle E\rangle = 32.6$ K and $18.3$ K.  The vertical lines indicate the fields at which quantum tunneling occurs for Mn$_{12}$-ac.  Note that the avalanche speed displays clear oscillations as a function of magnetic field, with higher values on-resonance than off-resonance due to quantum tunneling.  This can also be seen in Fig. \ref{speedsVsWf}.}

\label{fixedE}
\end{center}
\end{figure}

\begin{figure}[htbp]
\begin{center}
  \includegraphics[width=3in]{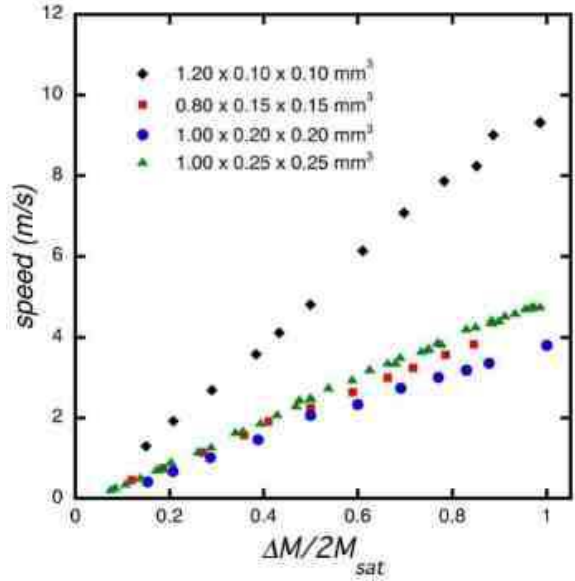}
\caption{Speed of propagation of Class II avalanches triggered in four different crystals at $\mu_0H_z = 2.5$ T for various initial magnetizations.}
\label{four Crystals}
\end{center}
\end{figure}

\begin{figure}[htbp]
\includegraphics[width=3in]{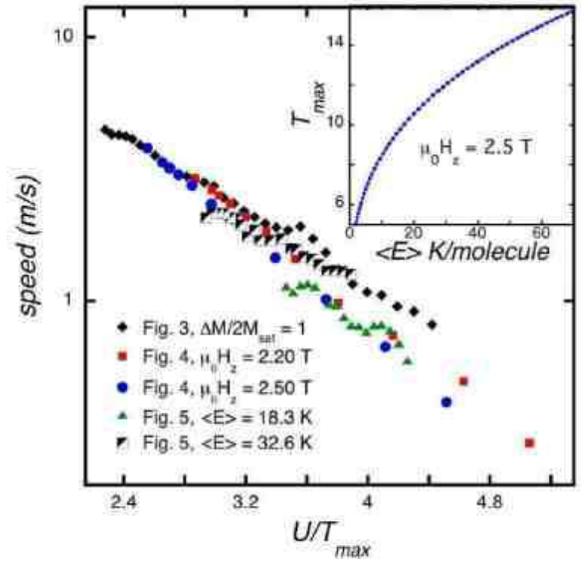}
\caption{Avalanche speed vs. the ratio $U/T_{max}$, where $U$ is the calculated barrier and $T_{max}$ is the maximum flame temperature.  The inset shows $T_{max}$ calculated from the heat capacity at $2.5$ T as a function of $\langle E \rangle$.}
\label{speedsVsWf}
\end{figure}

\begin{figure*}[htbp]
\includegraphics[height=3.25in]{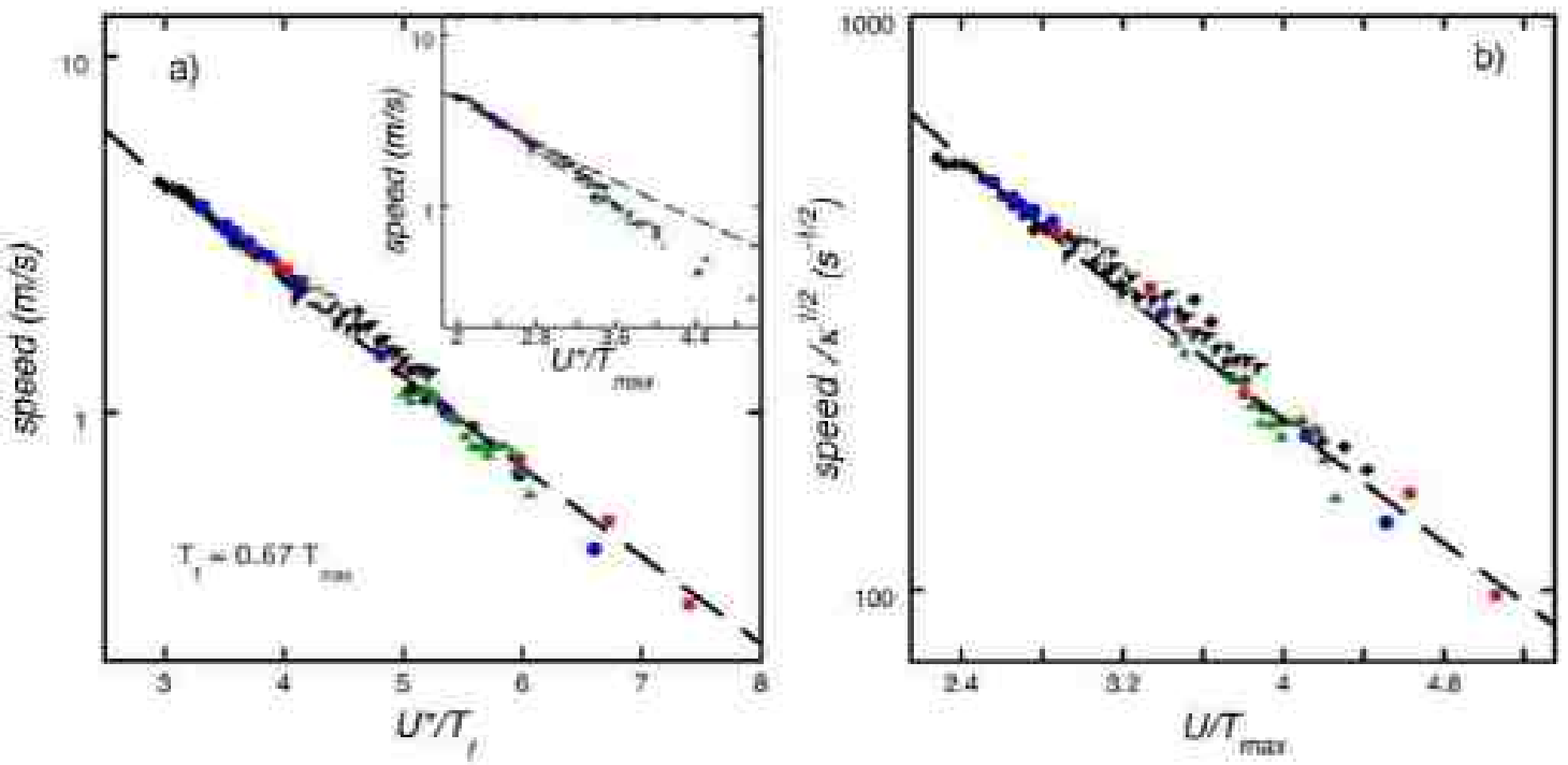}
\caption{(a) Avalanche speed vs. the ratio $(U^*/T_f)$, where the scaled barrier $U^* = (1-\alpha \frac{\Delta M}{2M_s})U$ with $\alpha = 0.13$, and the flame temperature $T_f = 0.67\times T_{max}$ with $\kappa = 1.2 \times 10^{-5}$ m$^2$/s.  The inset illustrates that a data collapse  can be obtained by adjusting $U^*$, (here $\kappa = 2.4 \times 10^{-6}$ m$^2$/s); however, numerical agreement with the measured data requires the additional  scaling parameter $T_f$.  (b)  Data collapse obtained using $\alpha$ calculated from the measured transverse field during the avalanche, setting $T=T_{max}$, and allowing the thermal diffusivity to vary with temperature.   The fit yields $\kappa = (1.8 \times 10^{-8})$  T$^{2.9}$ m$^2$/s.}
\label{fits}
\end{figure*}

Figure \ref{fixedM} shows the avalanche speed for class I avalanches triggered at various fields, all with the same initial metastable magnetization, $\Delta M/2M_{sat} = 1$.  The speeds of the avalanches increase as the field is increased.  This is expected since increasing the field both lowers the barrier and increases $\langle E \rangle$.  There are local maxima at fields corresponding to the tunneling resonances, which are denoted with vertical dotted lines.  This tunneling enhancement of the avalanche speed is consistent with previously reported results \cite{Alberto1, Alberto2, McHugh}.  It is also an indication that the flame temperatures are low enough to preserve the effective Hamiltonian, Eq. \ref{Hamiltonian}, and the rigid spin ($S=10$) approximation.  

Figure \ref{fixedU} shows the speed for class II avalanches triggered at fixed values of the external field.  In particular, data are shown for $\mu_0H_z = 2.50$ T and $\mu_0H_z = 2.20$ T.  By keeping the field fixed and varying the initial magnetization, only $\langle E \rangle$ varies while $U$ remains approximately fixed.  As $\Delta M/2M_{sat}$ decreases, so does the speed.

Figure \ref{fixedE} shows the speed for class III avalanches triggered at various external fields.  The initial magnetizations were also varied such that $\langle E \rangle$ remained approximately constant.  Presumably, the flame temperature is also nearly constant for all avalanches so prepared.  Therefore, the variation in avalanche speeds should be due to variation in the field-dependent barrier $U$.  Again, the vertical dotted lines drawn on Fig. \ref{fixedE} denote the location of the resonant fields for Mn$_{12}$-ac.  

Figure \ref{four Crystals} compares the class II avalanche speeds for all four crystals.  The speed of the avalanche varies considerably from sample to sample; however, any dependence on the sample dimensions is not immediately obvious.

\section{\label{Comparison Section}Comparison with Theory and Discussion.}

Garanin and Chudnovsky \cite{Garanin} developed a comprehensive theory of magnetic deflagration describing the ignition and propagation of the deflagration front.  For a planar front, the avalanche speed is given over a broad range of parameters by the simple approximate expression \cite{footI}:

\begin{eqnarray}
v = \sqrt{\frac{3\kappa T_f \Gamma(B, T_f)}{U(B)}},
\label{EOM}
\end{eqnarray}
where $T_f$ is the temperature at the front (flame temperature), $\kappa$ is the thermal diffusivity, $U(B)$ is the field-dependent barrier in units of Kelvin, and $\Gamma(B, T_f)$ is the relaxation rate of the metastable spins (Eq. \ref{Arrhenius}).  

Ref. \cite{dipole} established the importance of the dipolar fields in Mn$_{12}$-ac, as a fully magnetized sample adds (or subtracts) $\mu_0 M_z = \pm 52$ mT to the applied external field, $\mu_0H_z$.  We account for the initial magnetization, $M_z$, when calculating the various field dependent quantities using $B_z = \mu_0(H_z + M_z)$, where $-52$ mT $\leq M_z \leq 52$ mT (see Fig. \ref{MvsH}).  The barrier $U(B_z=\mu_0H_z + \mu_0M_z)$ is calculated from the effective spin Hamiltonian (Eq. \ref{Hamiltonian}). 

The average magnetic energy released by the relaxing spins (Eq. \ref{E}) leads to an increase in the temperature near the front.  Assuming no heat loss, the maximum possible temperature, $T_{max}$, can be calculated using the experimental heat capacity reported in Ref. \cite{Gomes}.  The heat capacity of Mn$_{12}$-ac depends on the magnetic field.  Therefore, we subtract the calculated zero-field spin (Schottky) contribution from the measured zero-field heat capacity from the data reported in Ref. \cite{Gomes}.  To this we add the calculated spin contribution at a specified field, $B_z$, for the total field dependent heat capacity $C_{tot}(B_z, T)$.  By equating the integral of this heat capacity to the average energy released per molecule, $\langle E\rangle$, we can calculate $T_{max}$,
\begin{eqnarray}
\langle E\rangle = \int_0^{T_{max}}C_{tot}(B_z, T)  dT.
\label{Heat capacity}
\end{eqnarray}
We assume the initial (ignition) temperature is much less than $T_{max}$.  This is a reasonable approximation, as the ignition temperatures for avalanches triggered above $1$T are below $1$ K (see ref. \cite{McHugh}) compared with values calculated for $T_{max}$ between 7 and 18 K (depending on $\langle E\rangle$).  $T_{max}$ (for $\mu_0 H_z = 2.5$ T) is shown as a function of $\langle E\rangle$ in the inset of Fig. \ref{speedsVsWf} (a).

We now proceed to compare our data with the theory of Garanin and Chudnovsky \cite{Garanin} as given by  Eq. \ref{EOM}.  If we assume that the thermal diffusivity $\kappa$ is a constant (or a weak function of temperature), then the speeds for all avalanches should lie on a single curve when plotted as a function of $U/T_{max}$.  Figure \ref{speedsVsWf} shows avalanche speeds for crystal A for the three different experimental protocols shown in Figs. \ref{fixedM}, \ref{fixedU}, and \ref{fixedE} plotted as a function of $U/T_{max}$.  Although the overall behavior for the three types of avalanches is similar, the data do not lie on one curve.

The deviations could arise from several factors.  (1) We have used an Arrhenius form for the magnetic relaxation; departures from Arrhenius law behavior are unlikely to be responsible for the deviations as it has been found experimentally to hold reasonably well in the range of temperature of our experiment \cite{Gamma_0, Gomes}.  (2) The thermal diffusivity is known to depend on temperature, while we have assumed it to be constant.  (3) The flame temperature may be lower than the value calculated from the specific heat, as some of the energy may escape the sample, or be distributed ahead of the front.  (4) The barrier $U$ may be reduced by the transverse component of the inhomogeneous field, $B_x$, established at the traveling front by the reversing spins (see inset to Fig. \ref{scopeData}).  The effects of $B_x$ can be included in the calculation of $U$ by including an additional Zeeman term $(-g\mu_BS_xB_x)$ in Eq. \ref{Hamiltonian}.  In addition, $B_x$ provides a symmetry-breaking term that increases the tunneling rate \cite{Friedman barrier, del Barco}.

We note that the deviations shown in Fig. \ref{speedsVsWf} are especially pronounced at high values of $U/T_{max}$.  In particular, the class I avalanches with $\Delta M/2M_{sat}=1$ (shown as filled diamonds) have the highest speeds.  This suggests that reduction of the potential barrier $U$ for large $\Delta M/2M_{sat}$ plays an important role.

It is unclear how to incorporate the effects of a spatially inhomogeneous transverse field component into the analytical theory of magnetic deflagration (Eq. \ref{EOM}).  Instead, we include the effects of $B_x$ on the relaxation rate by introducing an effective barrier:
\begin{eqnarray}
U^* \equiv \left(1-\alpha \frac{\Delta M}{2M_{sat}}\right)U,
\label{U*}
\end{eqnarray}
where $\alpha$ is determined empirically  \cite{footII}.  Although the scaling factor $\alpha\Delta M/2M_{sat}$ explicitly contains $M_z$ (thus appearing to be used twice), it is used here only to account for the size of $B_x$.  This is justified by our experiment, since the maximum value of $B_x$measured by the Hall sensors during an avalanche is found to be proportional to $\Delta M/2M_{sat}$.

The inset to Fig. \ref{fits} (a) demonstrates that a collapse onto a single curve is obtained for $\alpha = 0.13 \pm 0.01$.  However, the collapsed curve does not agree with the theory, shown by the dashed curve.  An additional step can bring them into line, as described below.

As pointed out earlier, due to possible heat loss through the edges of the crystal and/or heat diffusion ahead of the front, the flame temperature $T_f$ may well be less than $T_{max}$.  Assuming that $T_f$ is proportional to $T_{max}$, the constant of proportionality is deduced from fitting the data in the inset of Fig. \ref{fits} (a) with Eq. \ref{EOM}.  The diffusivity, still assumed to be temperature-independent, is also treated as a fitting parameter.  As shown in the main part of Fig. \ref{fits} (a), agreement with theory  is obtained for crystal A for $T_f = (0.67\pm 0.02)\times T_{max}$ and $\kappa = 1.2\times 10^{-5}$ m$^2$/s .  Using this analysis, all crystals show similar dependence on the barrier $U^*$ and the flame temperature $T_f$ and yield $T_f \approx 0.67 \times T_{max}$.  However, we find that the thermal diffusivity ranges from 1.2 $\times 10^{-5}$ to 8.7 $\times 10^{-5}$ m$^2$/s from crystal to crystal.

The fact that $T_f$ is the same fraction, $0.67\times T_{max}$, for all crystals is a puzzle.  The rate at which heat escapes the crystal during an avalanche must affect $T_f$.  The rate of heat loss is controlled primarily by the crystal cross section, surface roughness, and the thermal mounting conditions.  Variations in the mounting conditions inevitably occur (e.g., thickness of insulating grease), although every effort was made to use similar conditions from crystal to crystal.  There were no obvious visible differences in surface quality of the crystals.  The cross sections, however, were deliberately varied from $0.10\times0.10$ to $0.25\times0.25$ mm$^2$.  One expects that the crystals with smaller cross sections should lose more heat through the boundaries and should consequently have lower flame temperatures and, according to Eq. \ref{EOM}, smaller speeds.  Figure \ref{four Crystals} shows that the maximum speeds vary by approximately a factor of 2.5 from crystal to crystal, but without the expected dependence on cross section.   This implies that the widely different avalanche speeds in the four crystals (see Fig. 6) are unlikely to be due primarily to heat loss.  Instead, we suggest that the variation of the avalanche speeds are attributable to variations of $\kappa$.  The thermal diffusivity of dielectric crystals (like Mn$_{12}$-ac) at low temperatures is known to be strongly dependent on the defects, surface roughness, and dislocations in the crystal \cite{LowTemperaturePhysics}.

An additional puzzle is the large amount by which the potential barrier needs to be reduced to obtain a fit by the above analysis.  From our data, we deduce a  barrier $U^*$ that is 87\% of the classically calculated barrier, $U$.  A straightforward calculation implies that a transverse field of $\approx 0.4$ T  is required to reduce the barrier by that amount.  The largest $B_x$ field recorded by the Hall sensors during an avalanche is only $\sim 0.05$ T, an order of magnitude smaller.  Although it may contribute to it, the measured transverse field cannot by itself account for the large reduction of the barrier.

In the analysis presented above, the thermal diffusivity was assumed to be independent of temperature.  We now relax this condition.   We assume that the barrier $U^*$ is reduced by a much smaller amount corresponding to the measured transverse field, we set $T_f = T_{max}$, and we allow $\kappa$ to assume a temperature dependence that yields the best fit.  The result of this alternate fitting procedure, shown in Fig.  \ref{fits} (b), yields a collapse that is acceptable within the experimental uncertainties of the measurements.

Remarkably, the latter method of analysis yields a thermal diffusivity that $increases$ with increasing temperature approximately as $\kappa \propto T^3$.  This form seems quite unphysical, as the thermal diffusivity normally decreases as the temperature is raised.  We suggest that this unexpected behavior may be associated with a spin-phonon bottleneck.

A number of experiments have provided evidence that a spin-phonon bottleneck strongly affects the spin dynamics and energy relaxation at low temperatures in molecular magnets such as V$_{15}$ \cite{chiorescu}, Fe$_8$ \cite{jonathan,wernsdorfer} and Ni$_4$ \cite{kent}.  In this process, the Zeeman energy generated by the reversing spins does not find a sufficient number of phonon modes at low temperatures to allow direct energy relaxation and equilibration, so that thermal equilibrium is established slowly while the energy is ``bottlenecked'' in the spin system.  This bottleneck is lifted as the temperature increases, so that the number of available phonon modes increases and the energy is able to relax by direct spin-phonon processes.   The effect of the bottleneck can find expression within our analysis as either a departure from Arrhenius Law behavior (which we have assumed to be valid), or as an anomalous temperature dependence of the thermal diffusivity.   Within this scenario, $\kappa$ would not display the same temperature dependence when obtained by the standard method of measurement where one determines the time of propagation of a heat pulse, since in this case the energy is deposited into the phonon system directly.

\section{Conclusion}
We have presented the results of a thorough investigation in Mn$_{12}$-ac of the behavior of magnetic avalanches - the rapid reversal of magnetization that spreads through the crystal at subsonic speeds as a narrow interface between regions of opposite spin.  A controlled set of measurements in which some parameters were held fixed while others were varied provided systematic information, enabling a rigorous comparison with the  theory.


Two different methods were applied to fit data to the theory of magnetic deflagration of Garanin and Chudnovsky \cite{Garanin}.  In the first, we suggest that the internal transverse field produced by the avalanche front affects the speed of the front itself.  We model this effect with a reduced barrier, $U^*$, that varies with the size of the transverse field, $B_x$, produced by the avalanche front.  Assuming a constant, temperature-independent thermal diffusivity, a reduced barrier $U^*$ allows a collapse of all the data onto a single curve.  However, the transverse field measured at the front is not sufficiently strong to account for the large barrier reduction needed to obtain a good fit.  A barrier reduction of this magnitude, if correct, defies a simple classical analysis and may be a signal that quantum effects are important in the deflagration process for all values of $\mu_0H_z$, not just those associated with the tunneling resonances \cite{Friedman, Alberto1, Alberto2, McHugh}.

An alternative method of analysis that assumes a smaller barrier reduction commensurate with the measured values of transverse field yields a temperature dependent $\kappa \propto T^3$.  We speculate that this rather surprising temperature dependence may be real and due to a phonon bottleneck that becomes less effective as the temperature is raised.

To summarize, we find overall agreement between our measurements and the theory of magnetic deflagration of Garanin and Chudnovsky \cite{Garanin}.  However, detailed comparison yields either (A) a stronger reduction of the potential barrier than can be justified by the measured transverse fields; or (B) a thermal diffusivity that unexpectedly increases with increasing temperature, perhaps due to a phonon bottleneck; or (C) a combination of these (and possibly other) factors.  Further confirmation of the theory and a better understanding of the avalanche process, would be provided by a detailed theoretical analysis of bottleneck effects, and independent measurements of various parameters such as the flame temperature and the thermal diffusivity.

\section{Acknowledgements}
We are grateful to Eugene Chudnovsky, Dmitry Garanin, and Yosi Yeshurun for many helpful discussions.  We thank Hadas Shtrikman for providing the wafers from which the Hall sensors were fabricated.  S. M. thanks Liza McConnell for assistance with the data analysis.  This work was supported at City College by NSF grant DMR-00451605.  E. Z. acknowledges the support of the Israel Ministry of Science, Culture and Sports.  Support for G. C. was provided by NSF grant CHE-0414555.


\begin{thebibliography}{5}
\bibitem{minor complication}
{In order to avoid complications with the fast-relaxing minor species in Mn$_{12}$-acetate, the external field is chosen to be above $+1.25$ T, at which point, the minor species has already relaxed to the stable state and is uninvolved in the avalanche.}
\bibitem{!domain wall}
{Mn$_{12}$-ac is a superparamagnet at the temperatures greater than $\approx$ 3 K, which are produced during an avalanche. This precludes the possibility of a ferromagnetic domain wall.}
\bibitem{footI}
{The expression derived in Ref. \cite{Garanin} for the avalanche speed, $v$, contains a factor $\sqrt{4}$ rather than the $\sqrt{3}$ used here.  The factor $\sqrt{4}$ was derived assuming the heat capacity is of the the form $C\propto T^3$; however, a fit to the heat capacity of Ref. \cite{Gomes} yields $C\propto T^2$ in the temperature range of interest here.  Using the experimental temperature dependence in the derivation for $v$ gives the appropriate factor $\sqrt{3}$.

This expression for $v$ is correct for relatively slow relaxation rates, i.e., $U/T_f >> 1$ \cite{Garanin}, which we find is approximately satisfied by the avalanche data presented here (see Fig. \ref{speedsVsWf}).}
\bibitem{footII}
{It should be mentioned that the data can also be collapsed by scaling the temperature, $T^*=[1+\alpha(\Delta M/2M_s)]T_{max}$, rather than the barrier $U^*$.  We can find no physical justification for this, as the flame temperature is determined by the Zeeman energy released and not by the height of the barrier.}
\end{thebibliography}
\end{document}